\documentclass[12pt,twoside]{article}
\usepackage{xrb2007}
\usepackage{graphicx}

\newcommand{\nb}{Nuclear Bulge}
\newcommand{\xrs}{X-ray source}
\newcommand{\xrss}{X-ray sources}

\newcommand{\chandra}{\textit{Chandra}}

\newcommand{\spitzer}{\textit{Spitzer}}
\newcommand{\tmass}{2MASS}

\newcommand{\dr}{UKIDSS-GPS DR2}

\newcommand{\brg}{{\rm Br-}\ensuremath{\gamma}}
\newcommand{\nir}{near-infrared}

\begin{document}

\session{Faint Galactic XRB Populations}
\session{Faint XRBs and Galactic LMXBs}

\shortauthor{Gosling et al.}
\shorttitle{Nuclear Bulge X-ray source counterparts}

\title{Counterparts to the Nuclear Bulge X-ray source population}
\author{Andrew J. Gosling} \affil{Oxford Astrophysics, Department of
Physics, Denys Wilkinson Building, Keble Road, Oxford, OX1 3RH}
\author{Reba M. Bandyopadhyay} \affil{Department of Astronomy,
University of Florida, 210 Bryant Space Science Center, Gainesville,
FL 32611-2055} 
\author{Katherine M. Blundell} \affil{Oxford Astrophysics, Department
of Physics, Denys Wilkinson Building, Keble Road, Oxford, OX1 3RH}
\author{Phil Lucas} \affil{School of Physics, Astronomy and
Mathematics, University of Hertfordshire, College Lane, Hatfield,
Herts AL10 9AB}

\begin{abstract}

We present an initial matching of the source positions of the
\chandra\ \nb\ \xrss\ to the new UKIDSS-GPS \nir\ survey of the
\nb. This task is made difficult by the extremely crowded nature of
the region, despite this, we find candidate counterparts to $\sim
50\%$ of the \xrss. We show that detection in the $J$-band for a
candidate counterpart to an \xrs\ preferentially selects those
candidate counterparts in the foreground whereas candidate
counterparts with only detections in the $H$ and $K$-bands are more
likely to be \nb\ sources. We discuss the planned follow-up for these
candidate counterparts.

\end{abstract}

\section{\nb\ \xrss}

The \citet{wang02} and \citet{muno03} \chandra\ surveys of the \nb\
revealed a large population of weak, hard point-like \xrss\ that could
account for up to 10\% of the previously observed ``diffuse'' X-ray
emission from the \nb.  Numerous studies have tried to characterise
these newly discovered \xrss\ of the \nb\ surveys based on their X-ray
properties. \citet{pfah02} suggest that a large fraction may be
wind-accreting neutron stars with high mass
companions. \citet{muno03,muno06} and \citet{ruit06} propose that they
could be white dwarfs accreting from main sequence counterparts
(cataclysmic variables, polars and intermediate
polars). \citet{will03} and \citet{liu06} believe them to be neutron
stars with low mass companions and \citet{wu07} have speculated that
they could be isolated neutron stars and black holes accreting from
the interstellar medium. However, the weak nature of the sources means
that positive identification of the majority of the sources is
impossible based solely on the X-ray data. Identification of the
stellar counterparts to these \xrss\ we will allow the differentiation
between these possibilities.

As shown in \citet{band05}, to identify candidate counterparts to
\xrss\ in the Nuclear Bulge, very high resolutions imaging is required
to avoid issues of confusion due to the high stellar density of the
region. In addition to that, the extremely high, variable extinction
towards the Nuclear Bulge requires these observations to be performed
in the \nir\ or longer wavelengths. Previously, only the DENIS and
\tmass\ surveys had observed the \nb\ entirely in the \nir, however as
will be shown, their depth and resolution is insufficient for the
purposes of identifying the counterparts to the majority of the \nb\
\xrss.

\section{The UKIDSS-GPS}

The (United Kingdom Infrared Deep Sky Survey - Galactic Plane Survey)
UKIDSS-GPS is one of the five near infrared Public Legacy Surveys that
are being undertaken by the UKIDSS consortium, using the Wide Field
Camera on the United Kingdom Infrared Telescope
\citep{luca07}. Surveying the Northern Galactic Plane and the \nb\ in
$J$, $H$ and $K$ bands, integration times of 80s, 80s and 40s
respectively achieve median 5-$\sigma$ depths of $J=19.77$, $H=19.00$
and $K=18.05$ (Vega system) for the most recent Data Release 2
\citep[DR2; ][]{warr07}. The resolution of the UKIDSS-GPS observations
are 0.2\arcsec, an order of magnitude better than the \tmass\ survey,
with a similar positional accuracy of $\sim 0.1\arcsec$.

\begin{figure*}
  \begin{center}
    \includegraphics[width=0.99\textwidth]{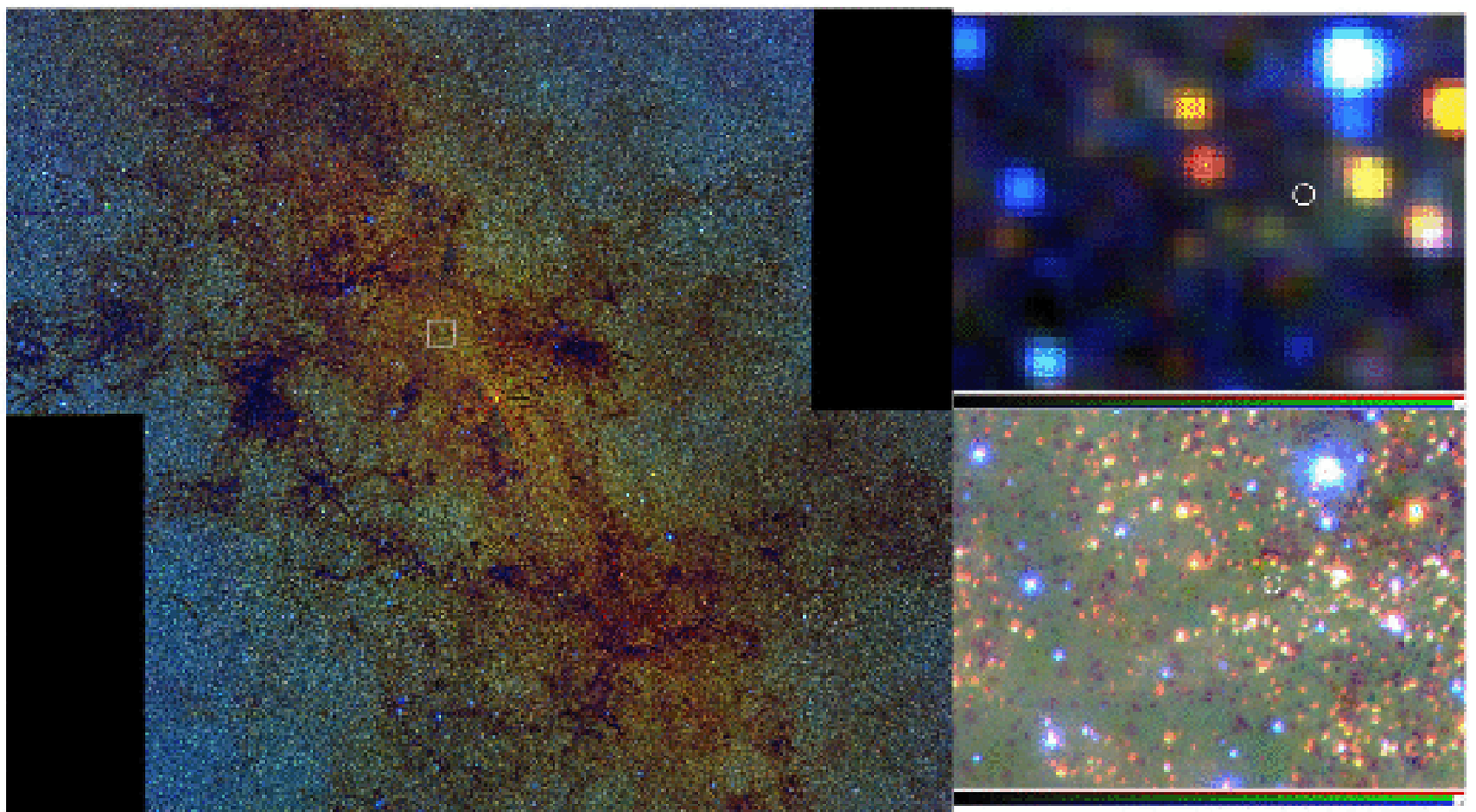}
    \caption[]{Near-infrared images of the \nb. {\it Left}: $\sim
    2\deg \times 2\deg$ false 3-colour image of the \nb\ in the \nir\
    (blue for $J$-band observations, green for $H$-band and red for
    $K$-band) containing $\sim 4,500,000$ point sources. The white box
    is the approximate location of the images displayed on the {\it
    right}. {\it Top-right}: old \tmass\ 3-colour image of the region
    indicated. {\it Bottom-right}: the equivalent image from the new
    UKIDSS-GPS survey demonstrating the vast improvement in both
    resolution and depth of the images. The white circle is the
    1\arcsec\ radius error circle of one of the \chandra\ \nb\ \xrss\
    demonstrating the scale of the images. }
    \label{f:nbimage}
  \end{center}
\end{figure*}

In the \nb, the crowding of this extremely dense stellar region
coupled with the high level of intervening extinction reduces the
completion limit of the survey to $J=18.5$, $H=16.0$ and $K=14.5$ with
detection limits of $J=20.5$, $H=19.5$ and $K=18.0$. Despite this, the
$\sim 2\deg \times 2\deg$ region shown in Figure \ref{f:nbimage}
contains $\sim 4,500,000$ point sources. As such it is the first
survey covering the entire \nb\ to have the resolution and depth to
allow the identification of candidate counterparts to the majority of
the \nb\ \xrss.

\section{\xrs\ counterparts}

We used the TOPCAT program to identify candidate counterparts to the
\xrss. We used a larger $0.3\arcsec$ positional error for the
UKIDSS-GPS \nb\ source positions, than the stated accuracy of
$0.1\arcsec$, to account for the confusion in the Nuclear Bulge caused
by the very high stellar density. The ``out-of-the-box'' \chandra\
positional error is $1\arcsec$, though this can be considerably larger
(up to $9\arcsec$) in some cases due to image distortion in the outer
regions of the \chandra\ array and/or to the very low S/N of the
faintest sources detected. For the purpose of this study, we discarded
all \xrss\ for which the X-ray positional errors were greater than
$3\arcsec$ radius; for error circles greater than this size, the
number of candidate counterparts is sufficiently high (5-30 stars) to
render meaningless any attempt at astrometric matching between the
\nir\ and X-ray catalogs. We used the $0.3\arcsec$ positional error
for the UKIDSS-GPS source positions, along with the derived positional
errors from the \citet{muno03} and \citet{muno06} \xrs\ catalogues to
set a maximum separation of $3.3\arcsec$ between and \xrs\ and a star
that can be considered a match.

\begin{table*}
  \begin{center}
    \caption[Number of X-ray source counterparts]{The numbers of
    candidate counterparts to the \citet{muno06} catalogue of X-ray
    sources from the \citet{wang02} \chandra\ observation of the
    Nuclear Bulge. The number of counterparts has been broken down in
    to \xrss\ with positional errors of $0\arcsec - 1\arcsec$,
    $1\arcsec - 2\arcsec$, $2\arcsec - 3\arcsec$ and the total numbers
    on the bottom row. }
      \vspace{5mm}
    \begin{tabular}{c|rrrr|r}
      \hline
      \hline
      X-ray error size & \multicolumn{5}{c}{Number of candidate counterparts} \\
      (arcsec) & 1 & 2 & 3 & $\geq 4$ & Total \\
      \hline
      $0\arcsec - 1\arcsec$ &  523 &  110 &   49 &   27 &  709 \\
      $1\arcsec - 2\arcsec$ &  600 &  153 &   58 &   43 &  854 \\
      $2\arcsec - 3\arcsec$ &  245 &   78 &   44 &   33 &  400 \\
      \hline
      $0\arcsec - 3\arcsec$ & 1368 &  341 &  151 &  103 & 1963 \\
      \hline
      \hline
    \end{tabular}
    \label{t:ncounterparts}
  \end{center}
\end{table*}

Of the 4256 \xrss\ in the \citet{muno03, muno06} catalogues, 3963
\xrss\ have positional errors of less than $3\arcsec$. From these, we
found 3076 candidate counterparts to 1963 of the \xrss. There are 1368
\xrss\ with only one candidate counterpart, 341 with two candidate
counterparts, 151 with three and 103 with 4 or more candidate
counterparts. Table \ref{t:ncounterparts} gives a breakdown of the
numbers of counterparts for \xrss\ with error circles of $0\arcsec -
1\arcsec$, $1\arcsec - 2\arcsec$ and $2\arcsec - 3\arcsec$. As can be
seen, for \xrss\ with positional errors of $0\arcsec - 2\arcsec$, the
majority of these have only one candidate counterpart, whereas
those with $1\arcsec - 2\arcsec$ errors have a slightly larger
proportion with 2 candidate counterparts. As might be expected, it is
the sources with the larger error circles, $2\arcsec - 3\arcsec$,
which have a much larger proportion with multiple candidate
counterparts.

\begin{figure*}
  \begin{center}
    \includegraphics[width=0.3\textwidth]{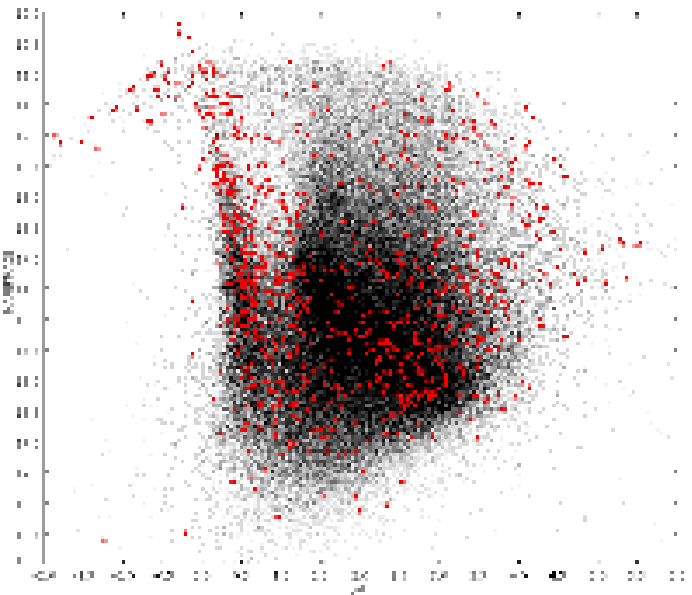}
    \includegraphics[width=0.3\textwidth]{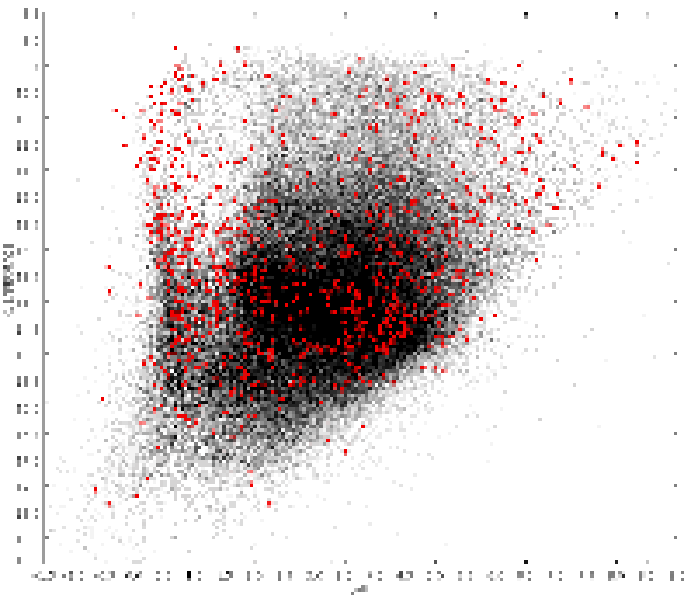}
    \includegraphics[width=0.3\textwidth]{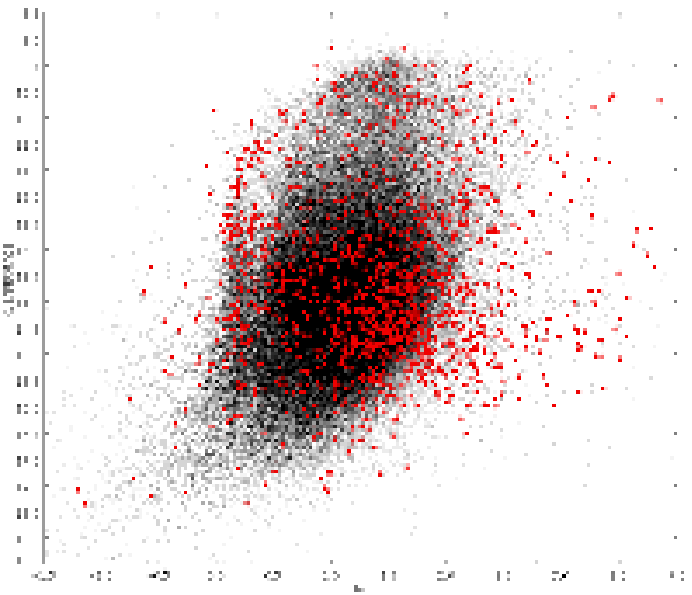}
    \caption[]{Colour magnitude diagrams (CMDs) of the candidate
      counterparts to the \chandra\ \xrss\ ({\it red} points) overlaid
      on a representative sample of the Nuclear Bulge stellar
      population ({\it black} points). From left to right the diagrams
      are $H\ vs\ J\!-\!H$, $K\ vs\ J\!-\!K$ and $K\ vs\ H\!-\!K$
      respectively. There is an increase in the number of candidate
      counterparts in the $K\ vs\ H\!-\!K$ CMD as the $H$ and $K$-band
      observations are less affected by the extinction towards the
      \nb.}
    \label{f:xrscolmag}
  \end{center}
\end{figure*}

Figure \ref{f:xrscolmag} shows colour magnitude diagrams (CMDs) of the
candidate counterparts compared to a representative sample of Nuclear
Bulge stars from the \dr\ catalogue. From the diagrams it can be seen
that the \xrss\ follow the general trend of the overall stellar
distribution but with some differences to their specific
distribution. For instance, the CMDs containing $J$-band data reveal a
high proportion of the candidate counterparts to be in the local,
main-sequence arm of the CMDs. The majority of the remaining candidate
counterparts in all three CMDs are consistent with red giants at a
Galactic Centre distance with variable levels of extinction (see
poster contribution by Gosling et al. for details of the \nb\
extinction).

Figure \ref{f:xrscolmag} also demonstrates the increase in candidate
counterparts that occurs when sources with photometry in the $H$ and
$K$-band is compared to those for which there is photometry in all 3
or $J$ and $H$-bands. There is a large amount of extinction towards
the \nb\ (see additional poster contribution of Gosling et al. for
further details), This extinction has a greater effect on shorter
wavelengths and so is most likely to obscure a source in the $J$ band.
This is also apparent in the increase in sources being predominately
those with large values of $H\!-\!K$ indicating high levels of
extinction.

\section{The next stages}

Identifying the candidate counterparts to the \nb\ \xrss\ is only the
first step in understanding this population. To better understand
them, it will be necessary to identify the true companion to each of
the \xrss\ where possible. In order to do this, we intend to search
for accretion signatures in the candidate counterparts. We are
undertaking a narrow-band Brackett-$\gamma$ (\brg, a known accretion
signature) imaging survey of the \nb\ to identify sources with \brg\
excess, and will also follow-up the candidate counterparts with \nir\
multi-object spectroscopy using the FLAMINGOS-2 instrument soon to be
commissioned on Gemini South. We will also compare our catalogue of
candidate counterparts to the \spitzer\ catalogue of point sources of
\citet{rami07} to increase our wavelength coverage of these sources
and to identify those with spectra indicating that they are likely to
be in accreting or X-ray emitting systems.

\section{Conclusions}

We have presented an initial matching of the source positions of the
\chandra\ \nb\ surveys of \citet{wang02} and \citet{muno03} to the new
UKIDSS-GPS \nir\ survey of the \nb. In doing so, we have identified
candidate counterparts to $\sim50\%$ of the \xrss\ in this extremely
crowded and heavily extincted region. We show that these candidate
counterparts are consistent with the overall stellar distribution of
observations towards the \nb\ although the relative proportions of the
populations are different.

Candidate counterparts that are observed in the $J$-band have a higher
tendency to be foreground sources and those observed only in the $H$
and $K$-bands are more likely to be \nb\ sources. This is most likely
to be an effect of the high levels of extremely spatially variable
extinction (see additional poster contribution by Gosling et al. for
further details) towards the \nb.

Further observations including narrow-band \brg\ imaging and \nir\
spectroscopy using the FLAMINGOS-2 instrument as well as comparison to
other datasets such as the \spitzer\ observations of the \nb\ will be
used to better identify the true companions to these \xrss\ and so
gain an understanding as to the nature of the \nb\ \xrs\ population.

\end{document}